\documentclass[runningheads]{llncs}
\usepackage[T1]{fontenc}
\usepackage{graphicx}
\usepackage{bbding}
\usepackage{hyperref}
\usepackage{booktabs}
\usepackage[misc]{ifsym}

\begin{document}
\title{\emph{ParliView:} An Experiment-Driven Platform for Parliamentary Transparency}

\titlerunning{\emph{ParliView:} An Experiment-Driven Platform for Parliamentary Transparency}

\author{
Eoghan Cunningham \inst{1}\Envelope \and
Rory Buckley\inst{2}       \and
Timea Balogh\inst{3}       \and
Louise Luxton\inst{3}      \and
Zoe Greene\inst{3}         \and
James Cross\inst{2}        \and 
Derek Greene\inst{1} 
}
\authorrunning{E. Cunningham et al.}
\tocauthor{Eoghan Cunningham, Rory Buckley, Timea Balogh, Louise Luxton, Zoe Greene, James Cross, Derek Greene}
\toctitle{\emph{ParliView:} An Experiment-Driven Platform for Parliamentary Transparency}
\institute{
  School of Computer Science, University College Dublin, Ireland\\
  \and
  School of Politics and International Relations, University College Dublin, Ireland\\
\email{\{firstname.surname\}@ucd.ie}
  \and 
  Department of Government and Public Policy, University of Strathclyde, Scotland \\
\email{\{firstname.surname\}@strath.ac.uk}\\
}

\maketitle  
\begin{abstract}
\textit{ParliView} is a modular, agentic retrieval-augmented generation platform for querying European Parliament records. It provides a natural language interface for citizens to explore parliamentary debates and activities, and a controlled experimentation environment in which different users can be exposed to systematically varied system behaviours. This enables rigorous study of how design choices, such as verbosity, interactivity, and source presentation, affect user trust, comprehension, and reliance. We describe the system architecture and demonstrate its two core interfaces.

\end{abstract}

\section{Introduction}

Parliamentary institutions generate vast quantities of records, such as debate transcripts, legislative documents, votes, and procedural minutes. While these records are nominally public, they are practically inaccessible to most citizens due to their volume and  complexity. These barriers contribute to a democratic deficit between citizens and the institutions that represent them: those who cannot see what representatives do are less likely to feel a stake in, or trust, them. Conversational AI systems offer a potential remedy by allowing users to pose natural language queries over parliamentary records, lowering the threshold for meaningful engagement with democratic processes \cite{rogiers2024kameraad}. However, deploying such systems responsibly raises important questions \cite{kim2025fostering}. To what extent do design choices, such as verbosity, interactivity, and source presentation, affect their trust, comprehension, and reliance on the system? To answer these questions, we require a robust and flexible system that facilitates experimentation.

We present \textit{ParliView},  a modular, agentic retrieval-augmented generation (RAG) platform in which a large language model (LLM) answers user queries grounded in European Parliament records. It provides both a conversational interface for citizens to explore  debates and activities, and a controlled experimentation environment in which different users can be exposed to systematically varied system behaviours. This enables rigorous study of how design choices affect user trust and comprehension, particularly the extent to which they induce over-reliance. In doing so, it provides the infrastructure needed to inform transparent, responsible, and effective deployment of AI tools in democratic contexts.

\section{System Architecture}

\noindent\textbf{Data.}
The \textit{ParliView} system draws on a range of European Parliament data. Unstructured sources, including debate transcripts, minutes, documents, and rules of procedure, are indexed in \textit{OpenSearch}\footnote{\url{https://opensearch.org/}} vector embedding databases. Structured data, such as Members of European Parliament (MEP) affiliations and procedural records, are stored in \textit{Aurora}\footnote{\url{https://aws.amazon.com/rds/aurora/}} SQL databases. The data ingestion pipeline is fully automated and updates daily. 

\begin{figure}[!h]
    \centering
    \includegraphics[width=0.75\linewidth]{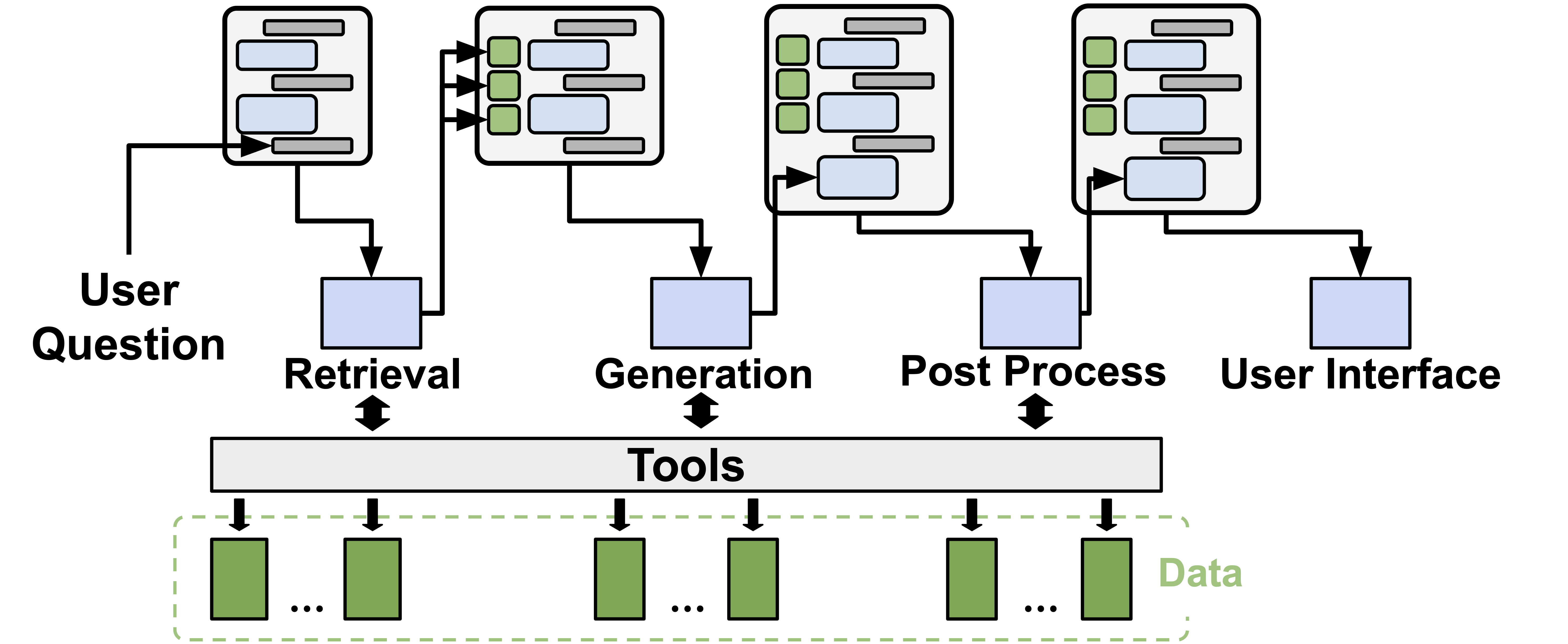}
    \vskip -0.7em
    \caption{\emph{ParliView} question-answering architecture, showing how user queries pass through a configurable pipeline of specialised components organised by function.}
    \label{fig:architecture}
\end{figure}

\vspace{0.2em}\noindent\textbf{Question Answering.}
The \textit{ParliView} system does not rely on a fixed question-answering pipeline. Instead, it is designed to allow researchers to reliably vary how the system behaves for different users and thus conduct systematic experimentation.  Figure~\ref{fig:architecture} presents an overview of the ParliView question-answering architecture. Distinct workflows can be composed from a library of specialised components organised by function.

\textit{Retrieval components} are agents that retrieve relevant information to ground the system's answer. They may vary in the databases they access, the retrieval mechanism used (e.g., semantic or hybrid search), and the querying behaviour of the agent. \textit{Generation components} are LLM-based agents (currently Claude Sonnet 4.6) tasked with answering the user's questions, and can be prompted to adjust response characteristics such as tone and verbosity. \textit{Post-processing components} are applied after an answer has been generated, and can reword responses, reformat citations, enforce guardrails, or redirect the query to further retrieval or generation components. \textit{User interface components} determine how information is presented to the user; by default, users receive transparent updates about the retrieval process and can explore all cited sources.





\begin{figure}[!t]
    \centering
    \includegraphics[width=0.85\linewidth]{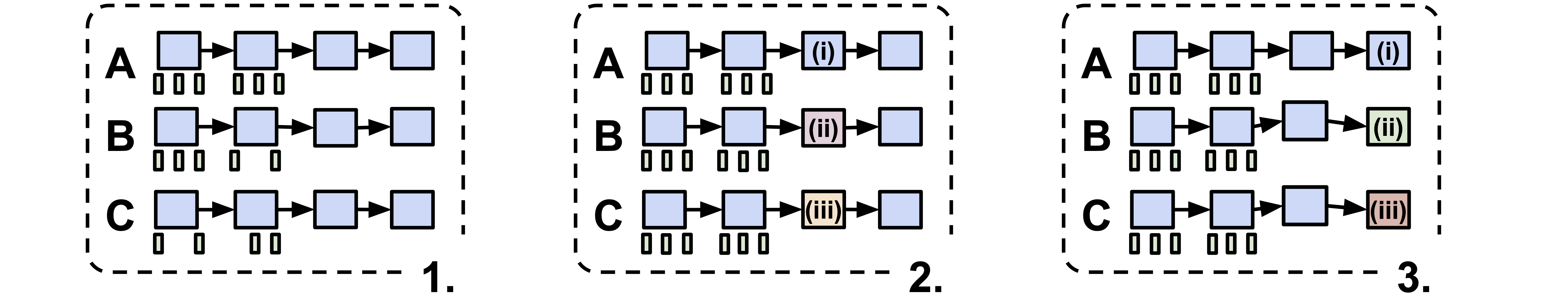}
        \vskip -0.8em
    \caption{Example experimental configuration with different treatment groups, each defined by a distinct DAG. Nodes may vary in the data and tools they access, the models and prompts they use, or the front-end features they present.}
    \label{fig:example_dags}
\end{figure}

Researchers construct experimental workflows by combining these components into Directed Acyclic Graphs (DAGs), where nodes represent agents and edges define conversation state flow. 
Each DAG represents a complete agentic workflow, as illustrated in Figure~\ref{fig:example_dags}. An experiment contains multiple treatment groups, each with its own DAG. 

\vspace{0.2em}
\noindent\textbf{Groups and Logging.}
To facilitate experimentation, \textit{ParliView} includes a group management and logging system. Participants are assigned to user groups, each associated with a distinct DAG, system prompt, and parameter set, ensuring that users' experiences vary according to the independent variables of a given experiment (e.g., transparency, interactivity, verbosity, tone, or use of jargon). The logging system records all user interactions, including questions asked, answers received, and sources explored. Alongside questionnaires, these logs serve as the dependent variables for a given experiment.


\section{Demonstration}

\begin{figure}[!t]
    \centering
    \includegraphics[width=0.9\linewidth]{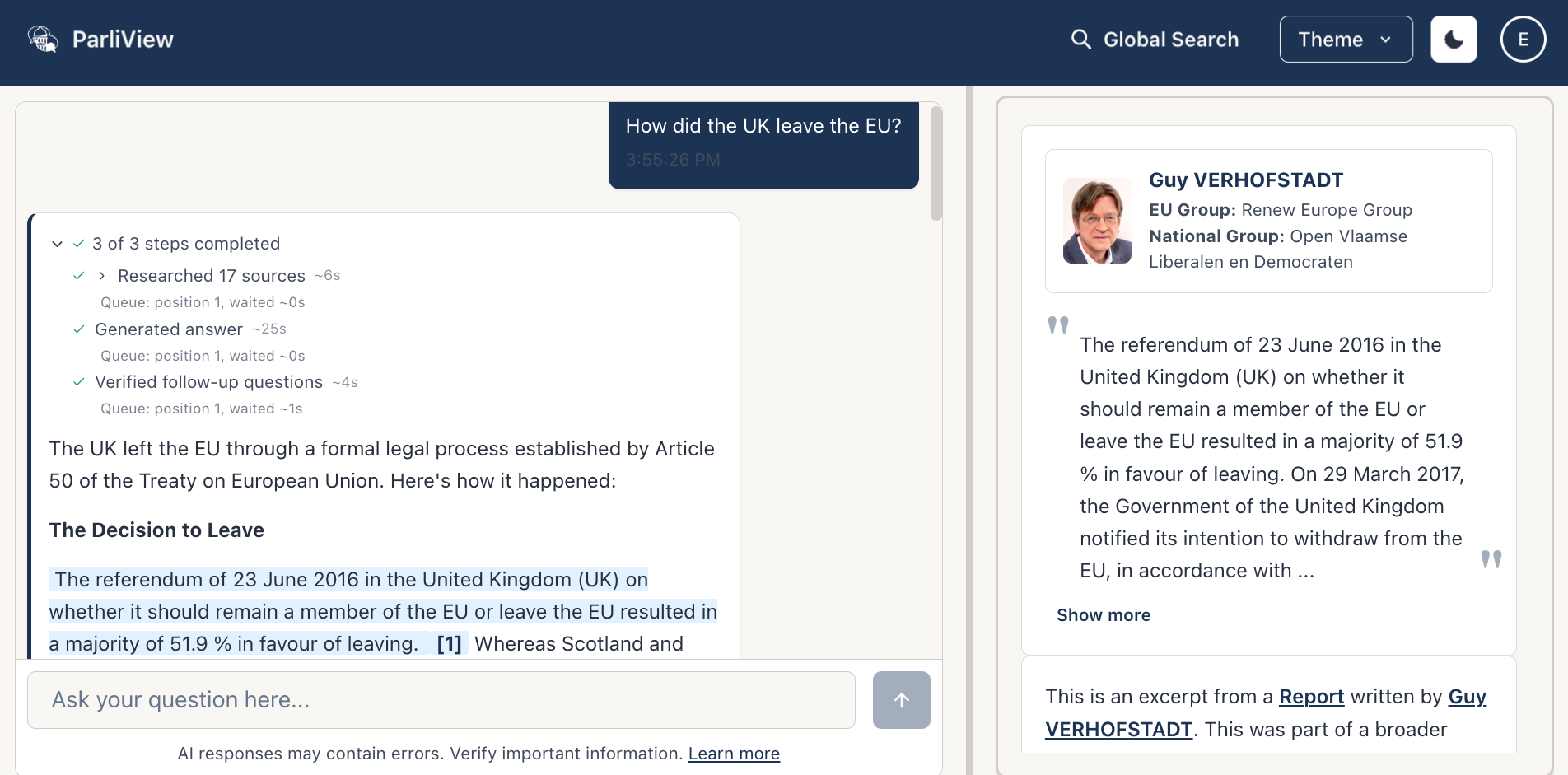}
    \vskip -0.7em
    \caption{The default citizen interface for \textit{ParliView}, showing transparent retrieval updates and access to cited sources.}
    \label{fig:citizen_view}
\end{figure}

\textit{ParliView} provides two key interfaces (see \url{www.parliview.org/ecml-demo}):

\vspace{0.4em}\noindent\textbf{Citizen Interface.} 
This interface allows users to explore parliamentary records through natural language questions (see Figure~\ref{fig:citizen_view}). By default, users receive transparent updates during answer generation and can access all cited sources in the sidebar. As outlined above, the system's flexible design means that this experience can be varied to suit a given hypothesis.

\begin{figure}[!t]
    \centering
    \includegraphics[width=0.9\linewidth]{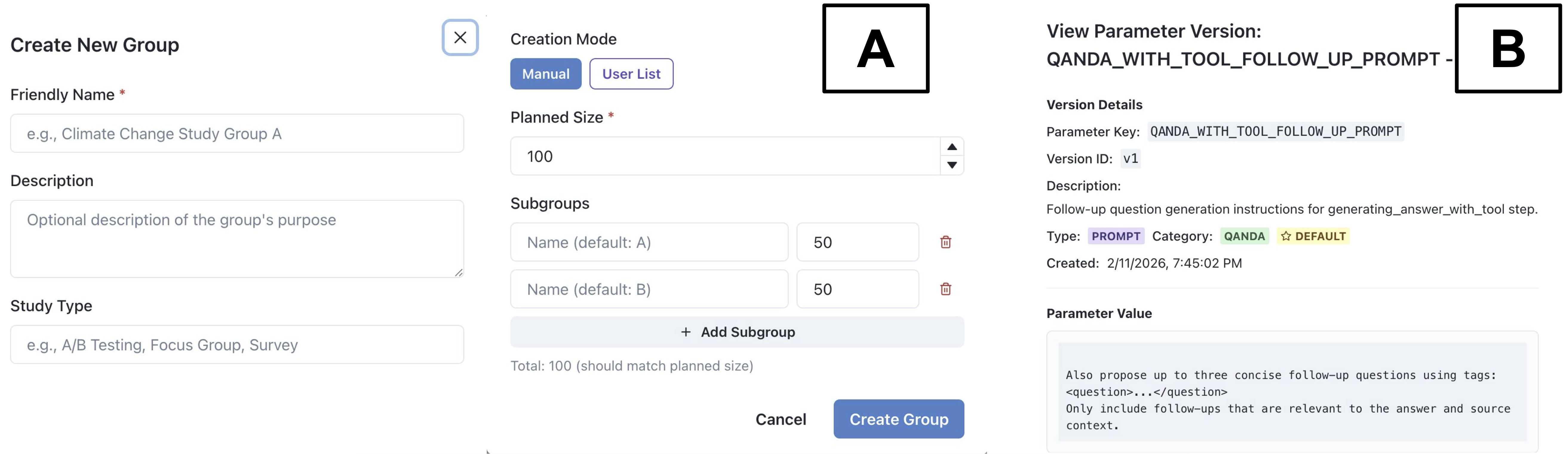}
    \vskip -0.7em
    \caption{The researcher interface for \textit{ParliView}. Administrators and researchers can (A) set up groups and specify sub-groups (e.g., control and treatment), and (B) view and override prompts and parameters.}
    \label{fig:researcher_view}
    \vskip-2em
\end{figure}

\vspace{0.2em}\noindent\textbf{Researcher Interface.} 
This interface allows researchers and administrators to define experiments, allocate users to groups (automatically or manually), and vary the question-answering workflow associated with each group (see Figure~\ref{fig:researcher_view}).

\section{Conclusions and Future Work}

\textit{ParliView} is a modular, agentic RAG platform for querying European Parliament records. The system's flexible architecture supports systematic experimentation on how design choices in AI-powered information systems affect user trust, comprehension, and agency, and the extent to which they induce over-reliance.

While \textit{ParliView} is designed primarily as a parliamentary transparency tool, its experimentation infrastructure generalises naturally to support wider research on user behaviour in LLM-based systems, such as how users form mental models of such systems, or how varying levels of transparency and interactivity affect long-term reliance.


%
%
%
%
%
%
\bibliographystyle{splncs04}
\bibliography{references.bib}

@inproceedings{rogiers2024kameraad,
author = {Rogiers, Alexander and others},
title = {{KamerRaad: Enhancing} Information Retrieval in {Belgian} National Politics Through Hierarchical Summarization and Conversational Interfaces},
year = {2024},
isbn = {978-3-031-70370-6},
publisher = {Springer-Verlag},
address = {Berlin, Heidelberg},
url = {https://doi.org/10.1007/978-3-031-70371-3_30},
booktitle = {Machine Learning and Knowledge Discovery in Databases. Research Track and Demo Track. ECML PKDD 2024},
numpages = {4},
location = {Vilnius, Lithuania}
}

@inproceedings{kim2025fostering,
author = {Kim, Sunnie S. Y. and others},
title = {Fostering Appropriate Reliance on Large Language Models: The Role of Explanations, Sources, and Inconsistencies},
year = {2025},
isbn = {9798400713941},
publisher = {ACM},
address = {New York, NY, USA},
doi = {10.1145/3706598.3714020},
booktitle = {Proceedings of the 2025 CHI Conference on Human Factors in Computing Systems},
articleno = {420},
numpages = {19},
location = {
},
series = {CHI '25}
}
\end{document}